\documentclass[lnicst]{svmultln}
\pdfoutput=1
\usepackage{epsfig,ifthen,endnotes,listings,subfigure,amsmath}
\usepackage{xspace}
\usepackage{algorithm}
\usepackage{algorithmic}

\newcommand{\eat}[1]{}

\newcommand{\honeytoken}{Mirage}

\usepackage{url}

   {%
      \end{minipage}%
        \end{center}}

\newenvironment{smitemize}%
  {\begin{list}{$\bullet$}%
     {\setlength{\parsep}{6pt}%
      \setlength{\topsep}{6pt}%
      \setlength{\itemsep}{6pt}}}%
  {\end{list}}

\makeatletter
\long\def\unmarkedfootnote#1{{\long\def\@makefntext##1{##1}\footnotetext{#1}}}
\makeatother

\begin{document}
\title{\honeytoken : Mitigating Illicit Inventorying in a RFID Enabled Retail Environment}
\author{Jonathan White and Nilanjan Banerjee}
\institute{University of Arkansas, Fayetteville \\ \{jlw09, nilanb\}@uark.edu}
\thispagestyle{empty}
\maketitle

\begin{abstract}
Given its low dollar and maintenance cost, RFID is poised to become the enabling technology for inventory control and supply chain management. However, as an outcome of its low cost, RFID based inventory control is susceptible to pernicious security and privacy threats. A deleterious attack on such a system is {\em corporate espionage}, where attackers through illicit inventorying infer sales and restocking trends for products. In this paper, we first present plausible aftermaths of corporate espionage using real data from online sources. Second, to mitigate corporate espionage in a retail store environment,  we present a simple low-cost system called {\honeytoken}. {\honeytoken} uses  additional programmable low cost passive RFID tags called honeytokens to inject noise in retail store inventorying. Using a simple history based algorithm that controls activation and de-activation of honeytokens, {\honeytoken} randomizes sales and restocking trends. We evaluate {\honeytoken} in a real warehouse environment using a commercial off-the-shelf Motorola MC9090 handheld RFID reader and over 450 Gen2 low cost RFID tags. We show that {\honeytoken} successfully flattens and randomizes sales and restocking trends while adding minimal cost to inventory control. \\ 


\end{abstract}

\section{Introduction}
\label{sec:intro}

Due to its low dollar and maintenance cost, flexibility, and portability, RFID is poised to become the enabling technology for inventory control and supply chain management. In fact, supply chain moguls such as Wal-Mart have already adopted RFID-based tags and readers for inventory control. Every item in the retail store is tagged using low cost passive RFID---each tag has an unique identifier and an antenna and minimal logic circuit to interact with a reader. Unfortunately, in the effort to keep the cost per tag minimal, the wireless communication between a reader and the tag is highly simplistic---making it  susceptible to pernicious privacy threats~\cite{Juels2005}.

\begin{figure}[t!]
   \centering
   \includegraphics[width=4 in]{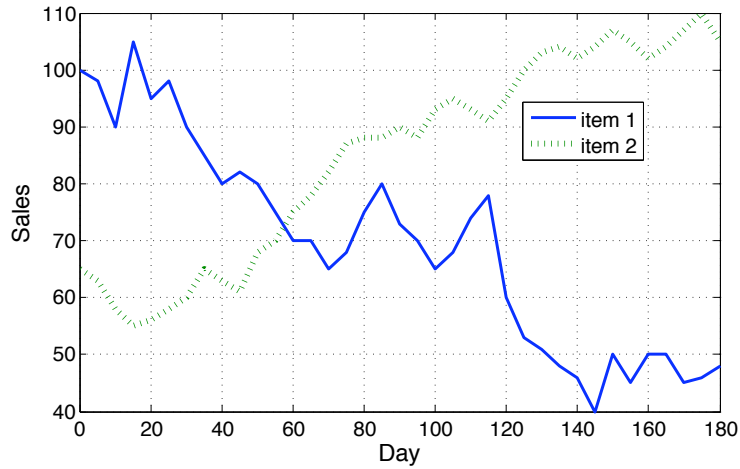} 
      \vspace*{5pt}
   \hrule
   \caption{The figure shows the trend in item sales for two competing video game products over a period of 6 months. Since the relative trend of product sales determine their stock prices, if an attacker observes and infers this trend, he can misuse this information to sabotage the revenue of a company. }
   \label{fig:example2}
\end{figure}

From a supply chain management perspective, a plausible deleterious attack is {\em corporate espionage}~\cite{RSA}. {\em Corporate espionage} refers to a threat where a set of individuals use sale and purchase trends to sabotage a product or a competitor. To illustrate the gravity of the attack, we present an example in Figure~\ref{fig:example2}. The figure shows the sales of two competing video game systems from June 2007 to January 2008. The graph depicts real data collected from a market research analyst group. Through illicit inventorying at representative locations, if a set of attackers dynamically infer the above trend, they could potentially misuse the information.  For example, since stock prices are determined by the relative item sales, stock traders could buy company stock when the sales are on the rise and sell stocks for a competing company whenever there is a dip---such planned attacks could kill the revenue of the manufacturing company.  

At the core of the above privacy threat lies the vulnerability in the enabling RFID technology. Given the low cost constraint on RFID tags (5-10 cents a piece), these devices are built with minimal logic circuit. Hence, they respond to a RFID reader wirelessly in plaintext---making them susceptible to eavesdropping attacks. Additionally, commercial off-the-shelf portable RFID readers can be used to read these tags. Hence, a group of individuals with body worn RFID readers can continuously inventory a retail store such as Wal-Mart to access trends in product purchase and sales. While cryptography on these tags is a simple solution to prevent illicit inventorying, the cost associated with the additional hardware circuit is a  huge deterrent. Our personal conversation with officials at Wal-Mart reveals that using cryptographic techniques~\cite{Vaudenay2006,Marti2007,JuelsSCN2005} that increase tag cost by an order of magnitude is economically infeasible, especially for low cost items---adding a dollar tag to a bottle of  milk worth two dollars is a 50\% overhead. Other techniques such as blocker tags~\cite{Rivest2003} that create privacy zones around a product have not seen the light of the day in corporate environments due to the associated inventory control overhead. For instance, a Wal-Mart official inventories millions of items multiple times a day---hence overheads of the order of minutes in reading tags is simply not an option. Additionally, retail store companies are reluctant to make major changes in their software and hardware infrastructure. This motivates the need for alternative low cost, low overhead solutions to mitigate {\em corporate espionage} that require minimal change to the in-place infrastructure.

In this paper, we present the design, implementation, and evaluation of {\honeytoken}, a RFID-honeytokens based system to minimize {\em corporate espionage}. Honeytokens are information security tools that operate on the theory of deception.  Honeytokens are related to honeypots, which are devices used to attract and trace attackers by presenting a seemingly worthwhile but unprotected computer.  Similarly, the central idea of {\honeytoken} is to add additional programmable tags (honeytokens) that adaptively inject {\em noise} in a retail environment to mask distinguishable sale and purchase trends. The key challenge is to spoof the attacker into believing that the added tokens are real while minimizing the overhead of reading additional tags. {\honeytoken} uses a simple history-based algorithm and a random ID generation algorithm to {\em in-situ} determine the number and corresponding IDs for active honeytokens.  By adaptively programming a subset of fake RFID tags, {\honeytoken} hides usable trends in inventorying. We evaluate {\honeytoken} in a real supply chain center using a Motorola MC9090 handheld RFID reader and over 450 Gen2 low cost RFID tags manufactured by Avery-Dennison.  Our results show that {\honeytoken} effectively masks several trends in supply chain management.
\section{Case for {\honeytoken}}
\label{sec:case}

In this section, we illustrate the aftermaths of {\em corporate espionage} using three different scenarios.
We present a convincing case for {\honeytoken} through analysis of sale and purchase trend data publicly available from marketing research forums. Note that this data is usually available at a half-yearly or yearly basis---hence, attackers cannot use this data directly to launch corporate espionage attacks.

\begin{figure*}[t!]
\begin{minipage}[t]{.5\textwidth}
\includegraphics[width=\textwidth]{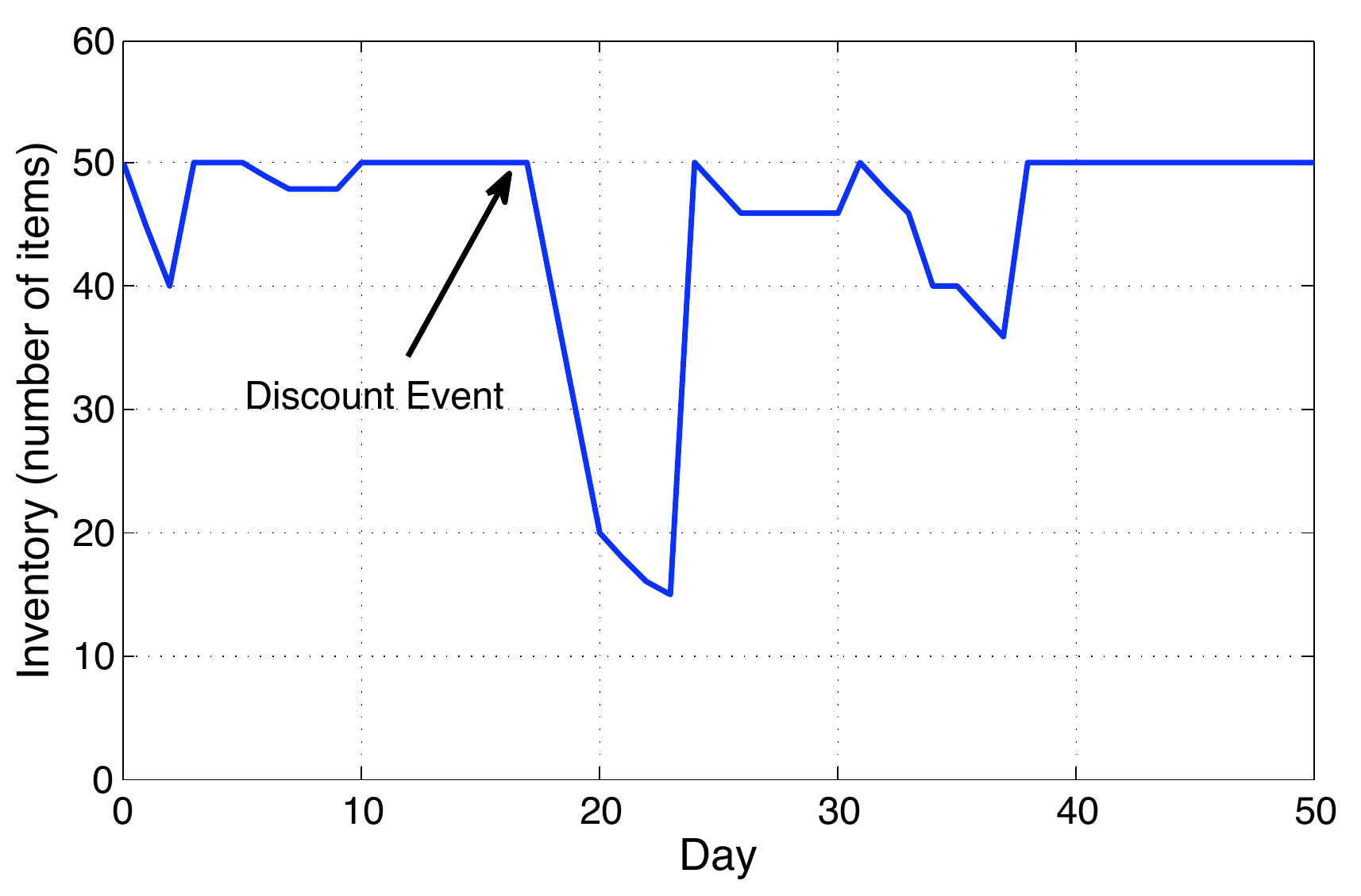}
   \vspace*{5pt}
   \hrule
\caption{The figure shows a scenario when a discount of items leads to heavy sales. Costumers who can study this trend can accurately predict how long they should procrastinate purchases. This could potentially lead to zero sales for an item.}
\label{fig:example1}
\end{minipage} \hspace{4pt}
\begin{minipage}[t]{.5\textwidth}
\includegraphics[width=\textwidth]{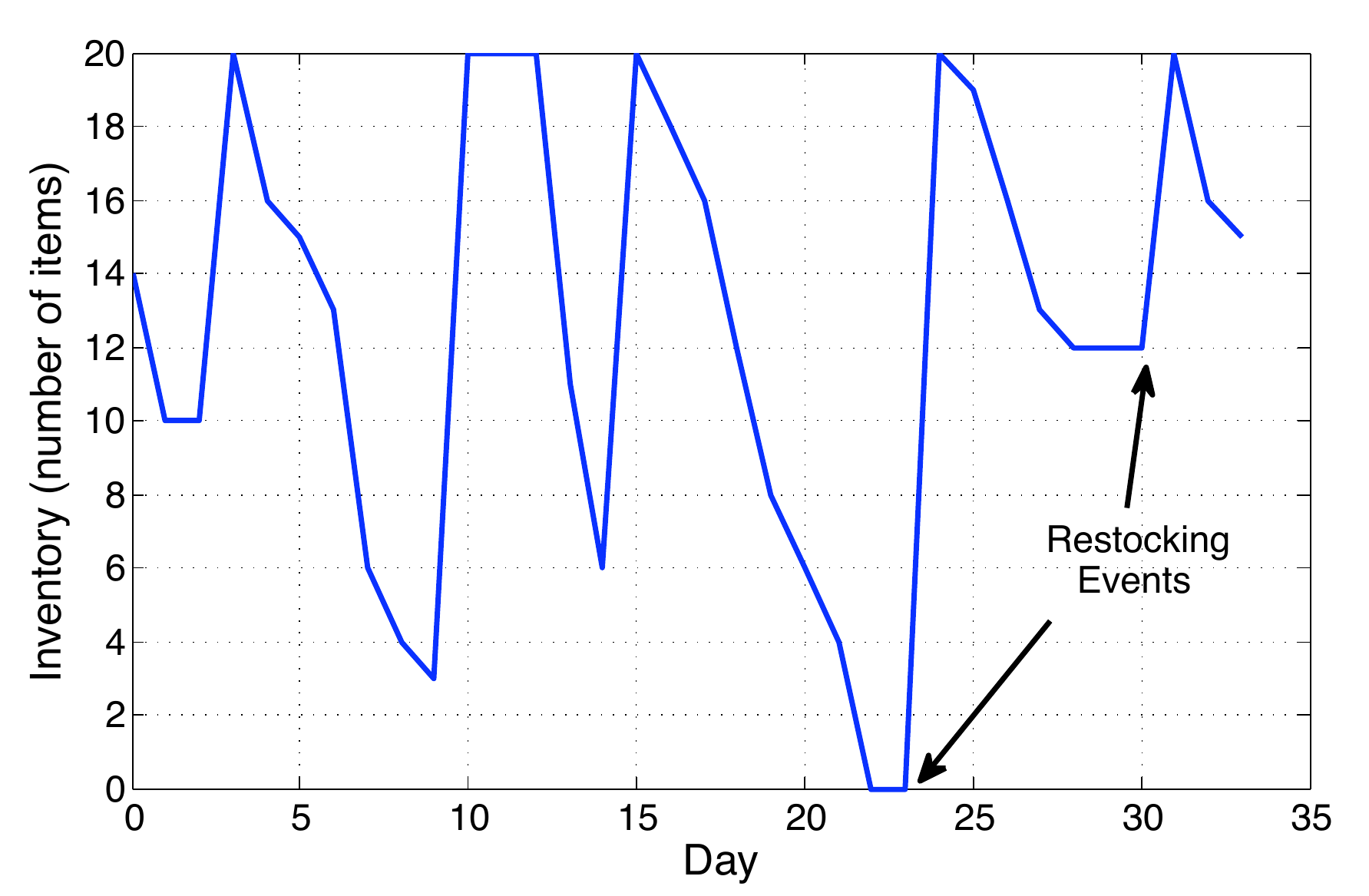}
   \vspace*{5pt}
   \hrule
\caption{The figure shows a trend that can be used to understand when restocking occurs in a company. This information could be used to launch a physical attack on transportation trucks and inventory houses. Similar attacks have taken place on banks.}
\label{fig:example3} \end{minipage}
\end{figure*}


\paragraph{Example 1:} Consider the sales trend of a popular product over a period of two months in Figure~\ref{fig:example1}. This data was publicly released by the US Census Bureau. The sales for the item is constant for a month, followed by a sudden spike when the items were sold at a discounted rate. This trend repeats itself after a month. If an attacker, through illicit inventorying at several outlets, deciphers this trend, he can infer accurate estimates of when the next discounted sale is likely going to take place---it seems like the manufacturer waits approximately a period of a month gauging sales and consequently announces a discount event. Such information can be used to lure customers to delay purchases until the next discount event, potentially sabotaging a company's revenues.


\paragraph{Example 2:} Figure~\ref{fig:example3} shows restocking trends for a popular
product. This data from January 2001 to July 2002 was collected from the US Census Bureau of Economics. The items are always restocked once their total falls below a threshold. An attacker can use this information to calculate the probability of a large restocking event given the number of items in stock. Physical attacks on transportation units carrying the product to retail stores could be apriori planned by leveraging this information. While other techniques such as ``bribing'', and ``hiring individuals'' could be used to plan such an attack, we believe that using illicit inventorying to do this is more ``covert''.

\paragraph{Example 3:} Figure~\ref{fig:example2} illustrates one of the more deleterious attacks using illicit inventorying (this data was collected from a NPD source). The figure shows the sales trend for two competing video game products. As outlined in \S\ref{sec:intro}, a third party company, through illicit inventorying at different  retail locations in the country can gain access to this trend. Consequently, the manufacturing company's stocks can be sabotaged.

\section{RFID-based Supply Chain Management}
\label{sec:rfid}

In the past,  UPC bar codes were used to physically scan inventory using a light-based scanner.  UPC tags require line-of-sight, which made inventorying tedious and error-prone. Since then, major retailers have chosen to migrate to RFID based systems.  Figure~\ref{fig:rfid} shows the three components of a RFID-based supply chain management system. Every item in a retail store such as Wal-Mart is tagged using passive RFID. A commercial off-the-shelf reader is used to read data from the tags---a 32-bits EPC (Electronic Product Code) code that identifies the object type, and an additional unique 32 bits product identifier~\cite{Juels2005,Juels2005EPC}. Data read by the scanner is offloaded to an Internet-resident database that keeps track of the items that are bought or sold at the retail store. The RFID reader is networked using Bluetooth or Wi-Fi.

\begin{figure}[t!]
   \centering
   \includegraphics[width=3 in]{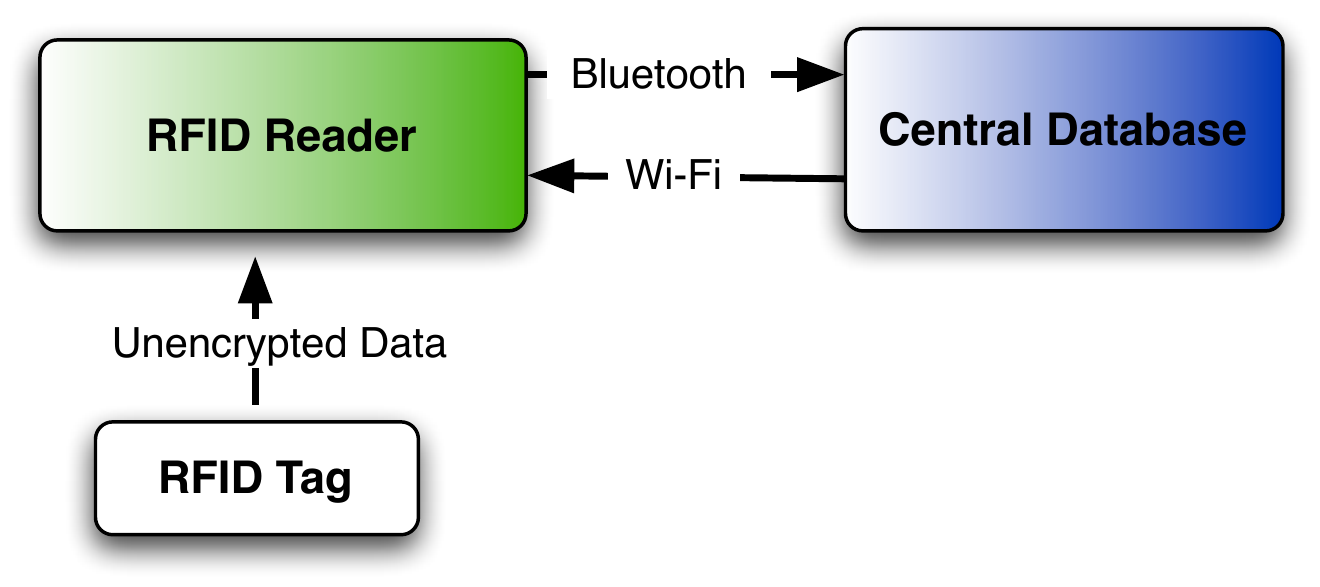} 
   \vspace*{5pt}
   \hrule
   \caption{The figure shows how inventorying is done in retail stores such as Wal-Mart. Off-the-shelf RFID readers are used to read unencrypted data from tags. This data is offloaded to an Internet-resident database using a Wi-Fi or Bluetooth connection. }
   \label{fig:rfid}
\end{figure}

The system is kept intentionally simplistic to minimize cost. Retail store management involves keeping track of several low-cost items. Adding encryption to RFID tags is not an economical option since it leads to an order of magnitude increase in dollar cost per tag. The result being, the system is prone to illicit inventorying attacks. Since the communication protocol used to offload data from a tag to a reader (Class1Gen2~\cite{ClassGen}) is  implemented on most commercial off-the-shelf readers, intruders with portable RFID readers can easily perform illicit inventories. If performed at several retail outlets simultaneously, attackers can infer trends in item sales, purchases, and restocking. This information can be used to launch several revenue attacks on manufacturers, as illustrated in \S\ref{sec:case}.

\section{{\honeytoken}}
\label{sec:mirage}

\begin{figure}[t!]
   \centering
   \includegraphics[width=3 in]{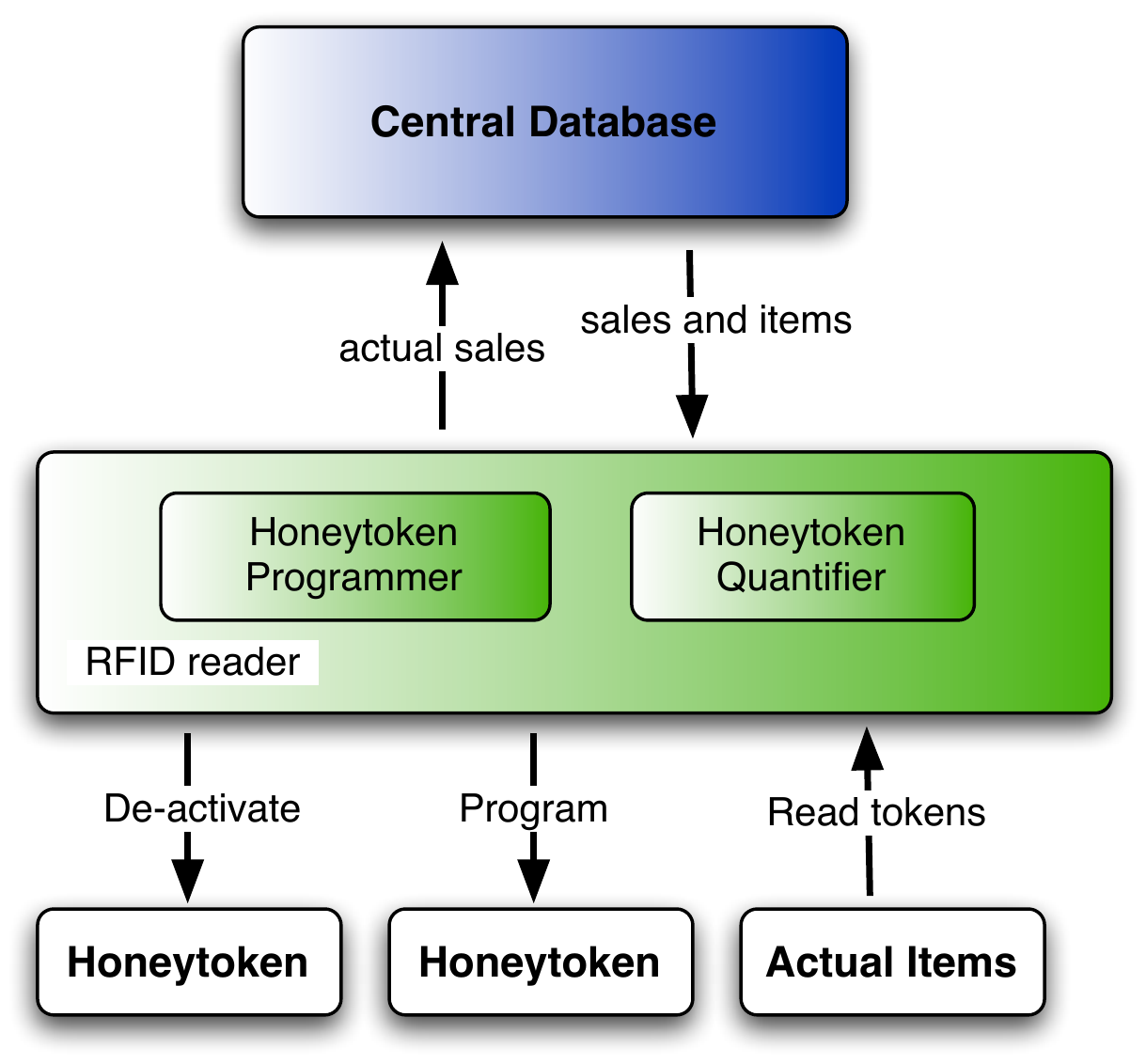} 
      \vspace*{5pt}
   \hrule
   \caption{The figure shows the general architecture of {\honeytoken}. The RFID reader reads data from tags (honeytokens and actual tags), downloads information from the central database on actual items in the inventory, calculates the number of honeytokens that needs to be programmed (to make the sales pattern look random), determines random IDs for the programmed honeytokens, and finally programs the tokens. The reader determines the actual tokens from the collection and sends an inventory on the real tags.  }
   \label{fig:rfid}
\end{figure}

Designing a {\em low overhead}, {\em simple} system that mitigates corporate espionage is challenging. First, the system should be {\em low cost}---hence encryption-based authentication is not an option. Second, the system should require minimal modification to the in-place infrastructure for easy adoption in a retail-store environment such as Wal-Mart. Third, and most importantly the scheme should mask both {\em sales} and {\em restocking} trends. To meet these goals, we design {\honeytoken}, a system that uses honeytokens to {\em randomize} sales and purchase trends in a retail environment. The central idea is to add a small number of additional programmable passive RFID tags that spoof as real items and inject {\em noise} into the retail store inventorying. However, the research challenge is to determine the {\em optimal} number of additional programmable tags per object type that should be added such that sales and restocking trends are {\em de-trended}. Consequently, when an attacker performs illicit inventorying, she sees random or flattened sales and restocking.

{\honeytoken} uses two inter-twined components (illustrated in Figure~\ref{fig:rfid}) to mitigate illicit inventorying---a ``Honeytoken quantifier'' and a ``Honeytoken programmer''. Both components are implemented on a programmable networked RFID reader. The ``Honeytoken quantifier'' determines the number of honeytokens that should be activated (or deactivated) such that both {\em sales} and {\em restocking} trends are randomized or flattened. It reads item tags and data from a secure back-end database to determine the actual sales and restocking numbers. It uses this information and a user-defined goal (of a random or flat distribution) in Algorithm~\ref{alg:algorithm} to determine the number of ``fake'' honeytokens to activate (or de-activate). In {\honeytoken}, the reader collects three important input parameters from back-end database tables: (1) the RFID tag IDs of honeytokens, (2) RFID tag IDs corresponding to real objects, and (3) the age of honeytokens. The ``Honeytoken programmer'' uses the identifiers for the actual and honeytoken tags to determine the IDs that new honeytokens should be programmed with.


\subsection{Honeytoken Quantifier}
\label{subsec:determine}

{\honeytoken} defines the notion of a {\em goal} that the honeytoken management system should meet. The goal of {\honeytoken} could be to portray a {\em flat} or a {\em random}  sales and restocking trend to an attacker. For instance, {\em flat} sales would always show constant sales to an individual performing illicit inventorying. To meet this {\em goal}, {\honeytoken} either activates or de-activates a set of honeytokens. Programming a honeytoken is equivalent to restocking one item on the shelf. De-activating a honeytoken amounts to a sales event from an attacker's perspective. The {\honeytoken} quantifier uses a simple history-based algorithm to calculate the number of honeytokens that should be programmed or de-activated.

{\em Flat distribution goal:} A flat distribution presents a {\em nearly-}constant sales and inventory trend to attackers. In our evaluation, we set the flat distribution goal of {\honeytoken} to 10\% more than the maximum sales in the previous month. However, this goal can be tuned to the availability of honeytokens at the warehouse. For instance, if only $x$ honeytokens are available and $y$ is maximum number of actual items on the shelf at any time, the goal could be set close to $x+y$.  {\honeytoken} uses the actual sales for an itemtype (difference in the number of items in the present and previous reader scan) and de-activates $|G  - S|$ honeytokens---$G$ is the flat distribution goal and $S$ is the number of actual items sold. Similarly, to meet a restocking goal of $G_i$, {\honeytoken} activates $|G_i - I|$ dead honeytokens, where $I$ is the actual number of items restocked. 

{\em Random distribution:} Another plausible goal is to depict a {\em random} distribution to the attacker.  To this end, {\honeytoken} uses a simple adaptation of the flat distribution approach. {\honeytoken} chooses a random sales or restocking goal for every time step---for example, this goal could be to show sales or restocking anywhere between the maximum and $100 + y$\% of the past maximum sales or restocking. Consequently,  using the same technique as above, {\honeytoken} deactivates or activates a certain number of honeytokens.  A primary challenge here is to assure that the random number generated on the RFID reader is {\em truly} random. We rigorously evaluate the  {\em randomness} of our output trend using autocorrelation measures in \S\ref{sec:evaluation}.

{\honeytoken} can also depict a flat or random {\em total inventory} trend. This is accomplished by a combination of sales and restocking goals. For instance, if the goal is to depict a flat inventory, {\honeytoken} tries to achieve a flat sales and restocking goal, the combination of which automatically leads to a flat inventory goal.

\begin{algorithm}[t!]
\caption{Honeytoken Quantifier Algorithm (at time $t_i$)}
\label{alg:algorithm}
\begin{algorithmic}
\STATE Input: $G$--- sales or restocking goal, $Av$---average age of items 
\LOOP
\STATE  For every Honeytoken $H$
\STATE Increment $Age(H)$ by one
\STATE {\bf if} $Age(H) > Av $  \texttt{deactivate token}
\ENDLOOP

\LOOP
\STATE $S(t_i)$ = sales between time $t_{i-1}$ and $t_{i}$
\STATE $R(t_i)$ = restocking between time $t_{i-1}$ and $t_{i}$
\STATE {\bf while} $S(t_i) < G$ \texttt{deactivate random token}
\STATE {\bf while} $R(t_i) < G$ \texttt{activate random unused token}
\ENDLOOP
\end{algorithmic}
\end{algorithm}

\subsection{Practical Issues}
\label{subsec:practical}

\begin{figure}[t!]
   \centering
   \includegraphics[width=3.5 in]{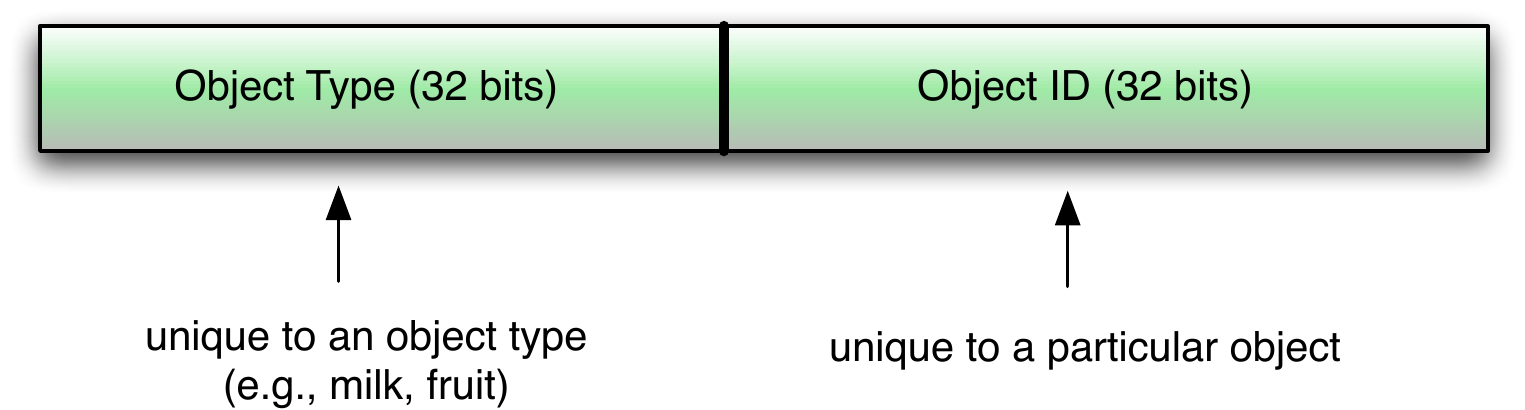} 
      \vspace*{5pt}
   \hrule
   \caption{The figure shows the breakdown of a honeytoken tag. The first 32 bits represent the type of the object (EPC code). The next 32 bits represent the exact object.}
   \label{fig:rfidtag}
\end{figure}

Using {\honeytoken} to circumvent illicit inventorying presents a number of practical issues. The most critical ones relate to the age of a honeytoken and the read overhead induced by the additional tags.

{\em Honeytoken age: } If the honeytoken IDs are not changed frequently, the attacker can isolate these from the real tags. For example, an attacker can watch the age distribution of a tag on the shelf---and use this distribution to distinguish the real tags from fake ones. To circumvent the above problem, {\honeytoken}  {\em ages} the deployed honeytokens. {\honeytoken} uses past history to determine the average age of a real tag on the shelf. If the age of a honeytoken is equal to the average age of real tags, the tag is reprogrammed with a random ID. Although this reprogramming accounts for one sales and one restocking event (potentially skewing results), it ensures that the attacker can not easily isolate an actual tag from a honeytoken. Further, it induces randomization in the inventorying process. Similarly, while deactivating honeytokens, tags whose age is equal or more than the average age of real tags are preferred. This ensures that the age of honeytoken tags follow a similar distribution as the  age of the actual tags---mitigating the threat of their exposure.  

{\em Read overhead: } Another practical problem is a result of the interference produced by the additional tokens. RFID tags respond to readers using a shared channel. Therefore, a large number of honeytokens can potentially lead to substantial collisions, incurring high read and reprogramming latencies. The solution to the above problem is to explore the trade-off between the following two parameters: ``how flat or random do we want the sales and restocking events to appear to an attacker'' and ``how much overhead is acceptable in terms of the number of honeytokens used''. Though in {\honeytoken}, we do not explicitly optimize for this overhead, in our evaluation we experiment with different number of honeytokens to determine what the ``optimal'' number should be such that the sales and restocking trend are randomized (or flattened) and the overhead associated with reading from and programming honeytokens is minimized.

Algorithm~\ref{alg:algorithm} is the core {\honeytoken} algorithm that determines which honeytoken tag should be programmed. It comprises of three major components. First, it keeps track of the {\em age} of a honeytoken. Using history, it determines the average age of an actual tag and uses it to decide whether a honeytoken should be reprogrammed. It tries to match a user-defined goal, and if depicted sales or restocking is more or less than the goal, a certain number of honeytokens are activated or deactivated. One of the primary goals of our system is {\em simplicity}. While more involved statistical techniques can be used to solve the problem, the trade-off lies in system complexity which directly effects the adoptability of the system. We show in our evaluation that the simple intuitive scheme used in {\honeytoken} suffices to solve the problem. Moreover, we believe that this would lead to easy adoption of {\honeytoken} by state-of-the-art retail stores such as Wal-Mart.

 \subsection{Honeytoken programmer}
 \label{sec:honeytoken}
 
 Activating or deactivating honeytokens is simple in {\honeytoken}. Every tag has a 32 bit identifier that is unique to an object. Another 32 bits are used to store information on the ``type'' of item (the schema of the tag is shown in Figure~\ref{fig:rfidtag}). When activating a tag, {\honeytoken} scans through all the tag IDs in the database and generates a new random ID---an ID that is not assigned to any other tag. The Item type field on the tag (first 32 bits) is kept exactly the same as the items on the shelf. When deactivating the tag, {\honeytoken} programs the item-type 32 bits with a random number. Hence, an attacker sees a different object type and automatically removes it from his list.

\section{Evaluation}
\label{sec:evaluation}

We evaluate {\honeytoken} with emphasis on two key questions.

\begin{smitemize}
\item Does {\honeytoken} obfuscate sales and restocking trends---does {\honeytoken} meet a flat or random trend goal ? 
\item What is the overhead in terms of reading and reprogramming honeytokens associated with {\honeytoken} ?
\end{smitemize}

\subsection{Experimental Setup and Methodology}
\begin{figure}[t!]
   \centering
   \includegraphics[width=4.5 in]{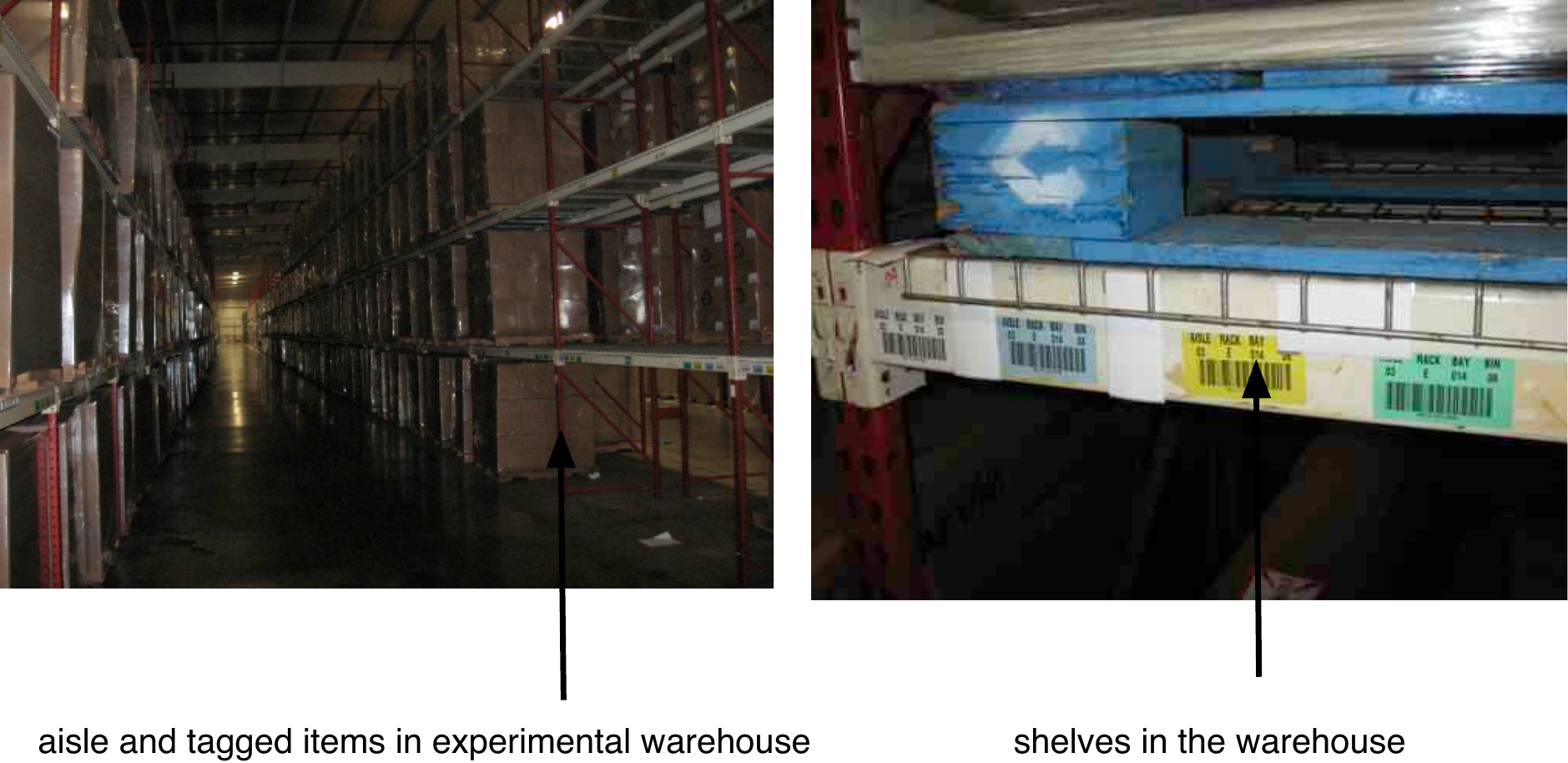} 
      \vspace*{5pt}
   \hrule
   \caption{The figure shows the warehouse where our experiments were conducted. The warehouse very closely mimics an actual Wal-Mart like stockroom.}
   \label{fig:Warehouse}
\end{figure}

\begin{figure}[t!]
   \centering
   \includegraphics[width=3 in]{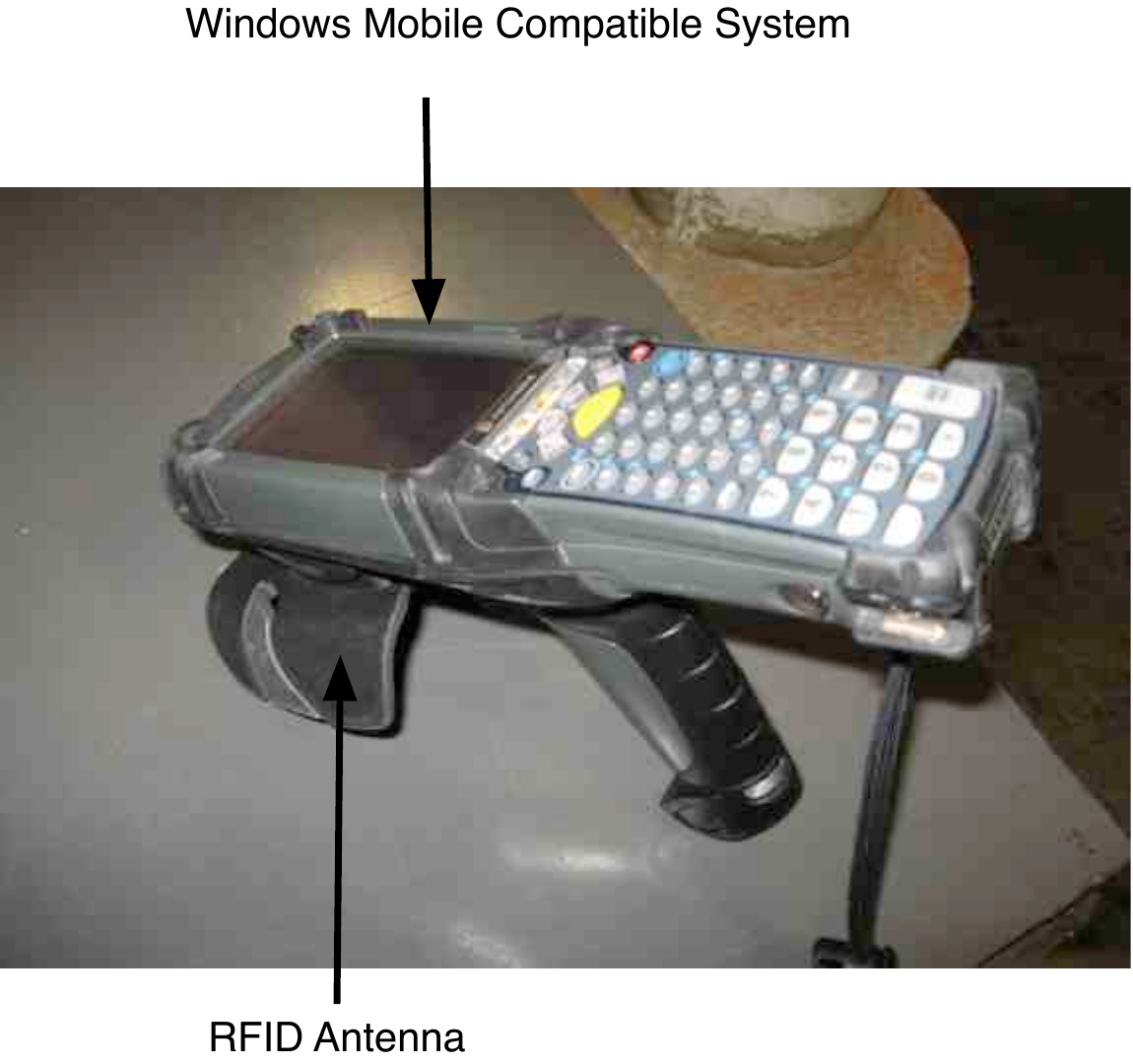} 
      \vspace*{5pt}
   \hrule
   \caption{The figure shows the Motorola RFID reader used in our experiments.}
   \label{fig:rfid_reader}
\end{figure}

For our evaluation, we use a real warehouse setup. The warehouse is an experimental center built specifically to run realistic tests on RFID-based inventorying. Figure~\ref{fig:Warehouse} is a picture of the aisle in the experimental warehouse where items are stored. These are real items sold at a Wal-Mart like store---therefore, practical issues such as reading delays due to metal and water still hold in our experiments. We use an off-the-shelf Windows-Mobile compatible Motorola MC9090 handheld RFID reader and over 450 Gen2 low cost RFID tags manufactured by Avery-Dennison in our experiments (see Figure~\ref{fig:rfid_reader}). The reader has a range of approximately 3.5 meters.

 A major challenge to realistic evaluation of {\honeytoken} is emulating real sales-restocking trends. Moreover, realistic evaluation using data presented in \S\ref{sec:case} would take years of experimentation. To expedite our evaluation, we use a ``time contraction'' approach. As an input to our experiments in the warehouse, we use the inventory from the three examples in \S\ref{sec:case}. Though the data in \S\ref{sec:case} was collected over a period of several months, the number of data points usually correspond to the number of scans done in the retail store. Instead of waiting for an entire day (sometimes several days) for a scan, we speed up the process by performing a scan, followed by removing/adding items to the shelf (corresponding to the actual data), and then immediately performing the second scan. Each scan corresponds to one time step in our experiment. For adding honeytokens, we follow a similar approach. We determine the honeytokens that need to be programmed and used in a particular time step and scatter them with the real items on the shelves. Consequently, we perform a read using the RFID reader. We present read and programming latency results in \S\ref{subsec:overhead}.

\label{subsec:detrend}
\begin{figure*}[t!]
\begin{minipage}[t]{.5\textwidth}
\includegraphics[width=\textwidth]{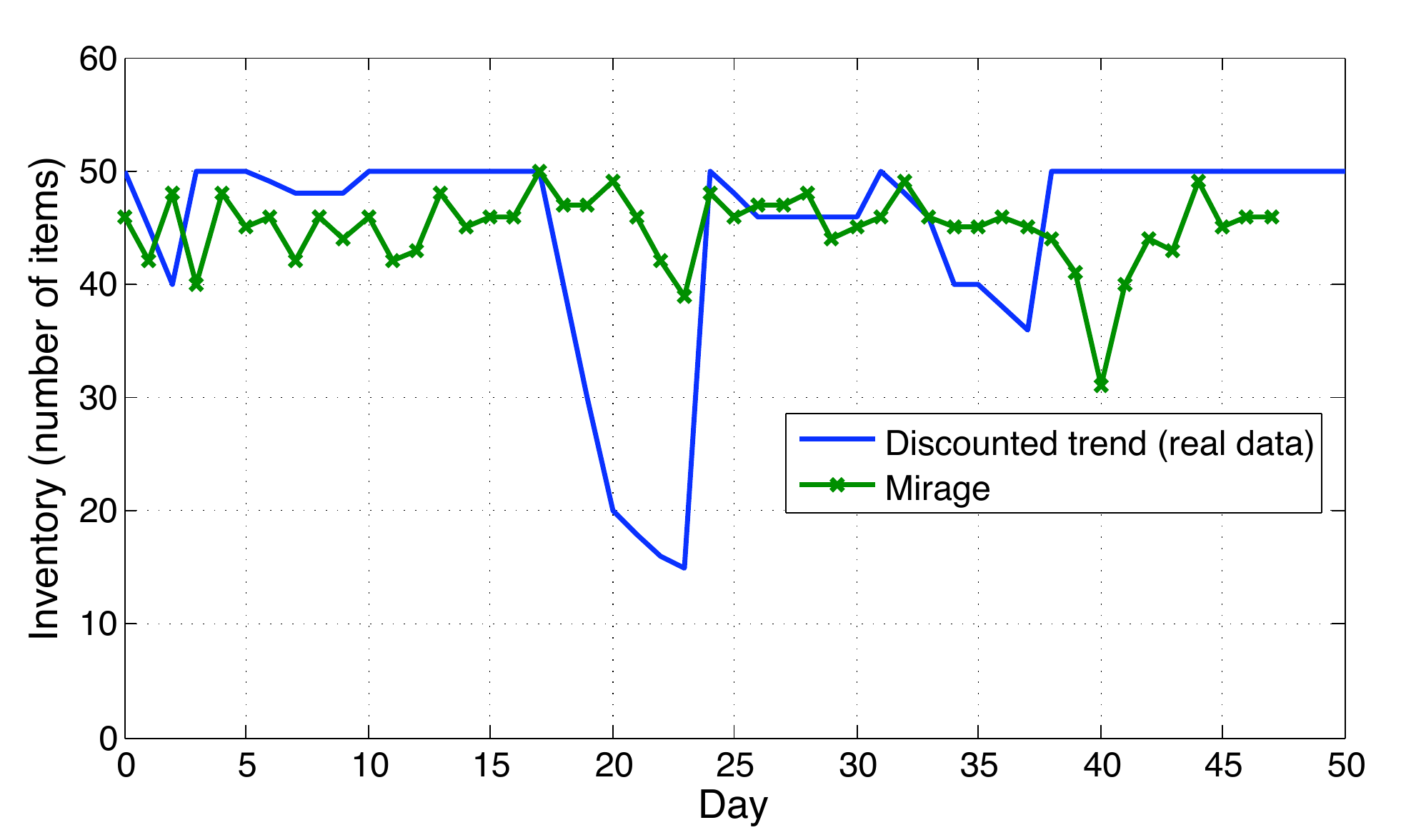}
   \vspace*{5pt}
   \hrule
\caption{The figure shows the flattened inventory trend for data in Figure~\ref{fig:example1}. Using  {\honeytoken}, the discount sales trend cannot be visually distinguished. }
\label{fig:flatdiscount}
\end{minipage} \hspace{4pt}
\begin{minipage}[t]{.5\textwidth}
\includegraphics[width=\textwidth]{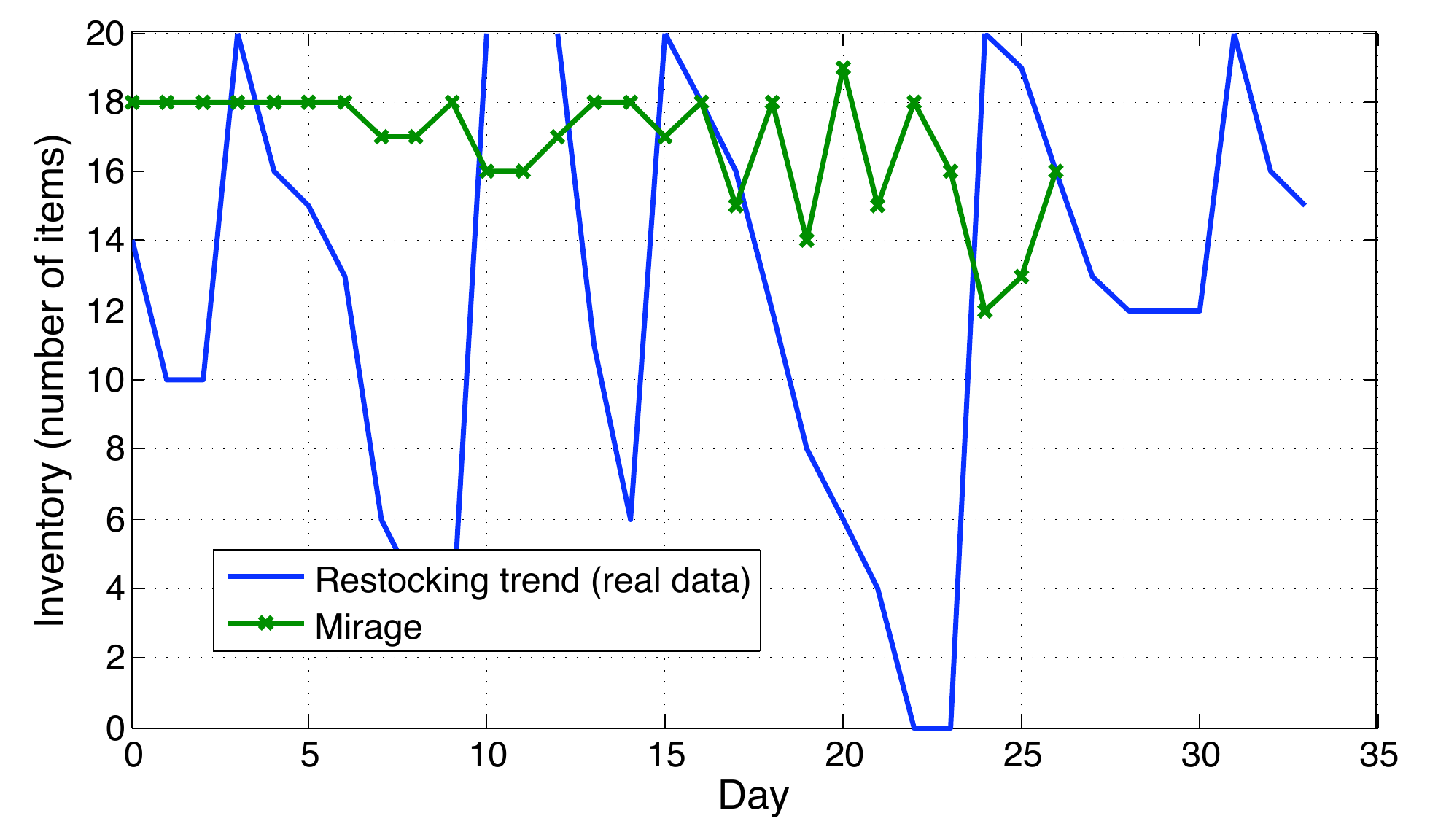}
   \vspace*{5pt}
   \hrule
\caption{The figure shows the flattened inventory trend for data in Figure~\ref{fig:example3}. {\honeytoken} successfully flattens the restocking events.}
\label{fig:flatrestocking} \end{minipage}
\end{figure*}

\subsection{De-trending sales and restocking}
Our first set of experiments evaluate de-trending restocking and discount sales for the first two scenarios in \S\ref{sec:case}. Figure~\ref{fig:flatdiscount} shows the results of using {\honeytoken} to flatten inventory trends corresponding to data in Figure~\ref{fig:example1} and Figure~\ref{fig:flatrestocking} presents the results of flattening inventory trend corresponding to data in Figure~\ref{fig:example3}. In the first example, the goal is to hide the discount event from an attacker and in the second example, the goal is to obfuscate the restocking events. 37 and 20 honeytokens were used for the two experiments respectively. From the results, we make two observations. First, {\honeytoken} uses honeytokens intelligently to produce a near-flat inventory to an attacker and successfully hides possible trends in the data. Second, we observe that in-spite of adding honeytokens, the number of tags read are less than in the actual data. This anomaly is explained by the read success probability of RFID tags. When a RFID reader tries to read tags in a retail-store environment, the probability of reading all the tags is less than 1 due to RF collisions, physical interference from metal or water, and tags that are at the edge of the reader range. In our warehouse setting, we found that 82\% of the tags were read on average during the 50 separate inventories, with the best inventory success rate of 89\%  and the worst inventory success rate of 57\%. We also note that these results correspond to the case where {\honeytoken} needs to provide a combination of flat sales and flat restocking such that the overall inventory trend is flat.

\begin{figure}[t!]
   \centering
   \includegraphics[width=4in]{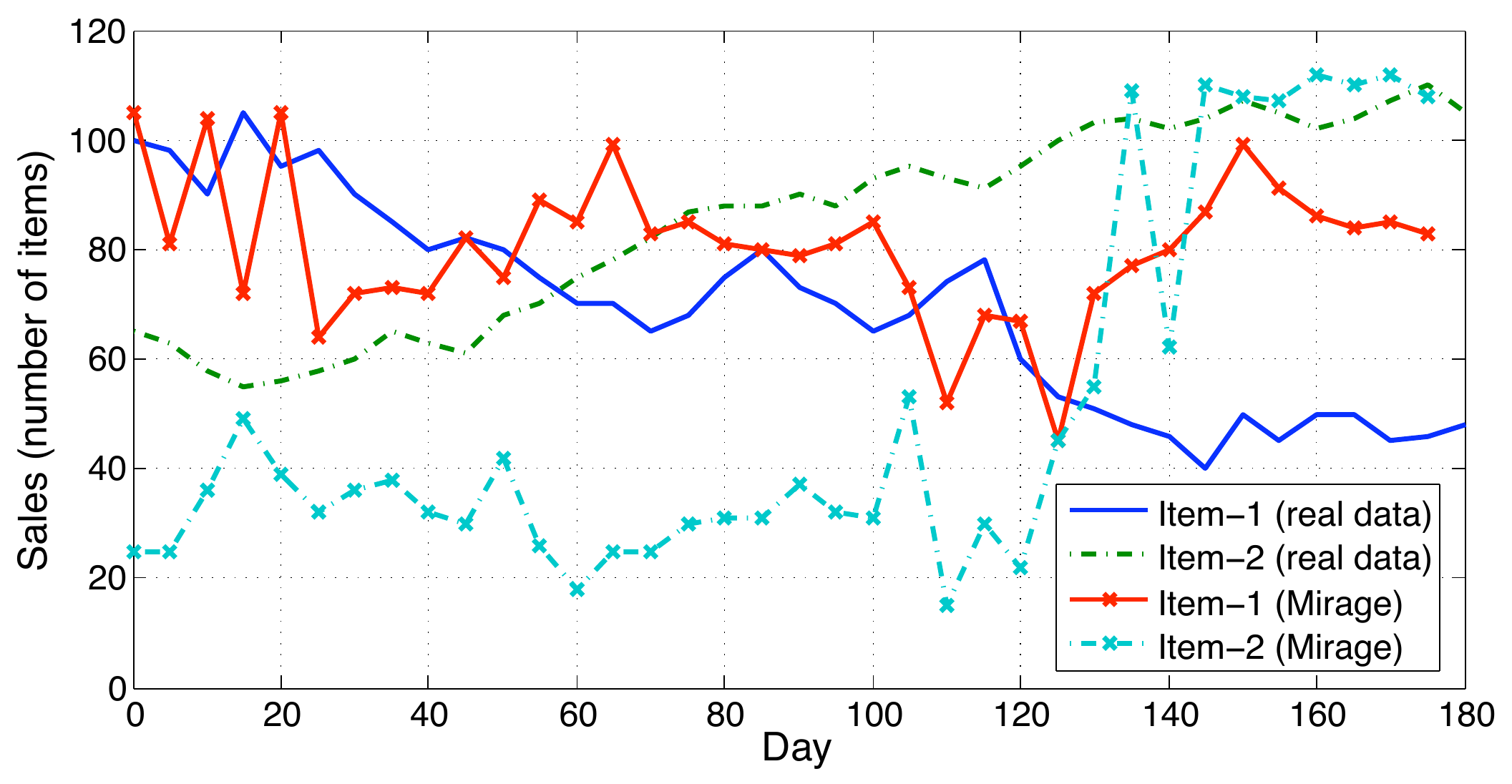} 
      \vspace*{5pt}
   \hrule
   \caption{The figure shows the results of the sales trend presented to an attacker by {\honeytoken}. The goal in this experiment was to obtain a flat sales distribution. {\honeytoken} does not obtain a flat distribution due to two factors, (1) the flat goal for a month is calculated as a function of the sales in the previous months, and (2) the success probability of the RFID reader to read a honeytoken tag in a warehouse is less than 100\%.}
   \label{fig:salesflat}
\end{figure}

Figure~\ref{fig:salesflat} shows the results of applying {\honeytoken} to the sales trends in Figure~\ref{fig:example2}. This experiment shows a limitation of trying to obtain a flat distribution for certain scenarios. We observe that for Item-2, the sales trend is still visually distinct. This is because {\honeytoken} uses the maximum sales from the previous month to determine the goal for the present month. As the sales pattern in Figure~\ref{fig:example2} has large variance, the resulting goals for each month oscillates, revealing the actual trend in some cases. Also every scan does not read all the RFID tags---in this experiment, 68\% of the total inventory was captured on average during the 36 separate scans, with 82\% being captured in the best case and 33\% captured in the worst case.

\begin{figure}[t!]
   \centering
   \includegraphics[width=4in]{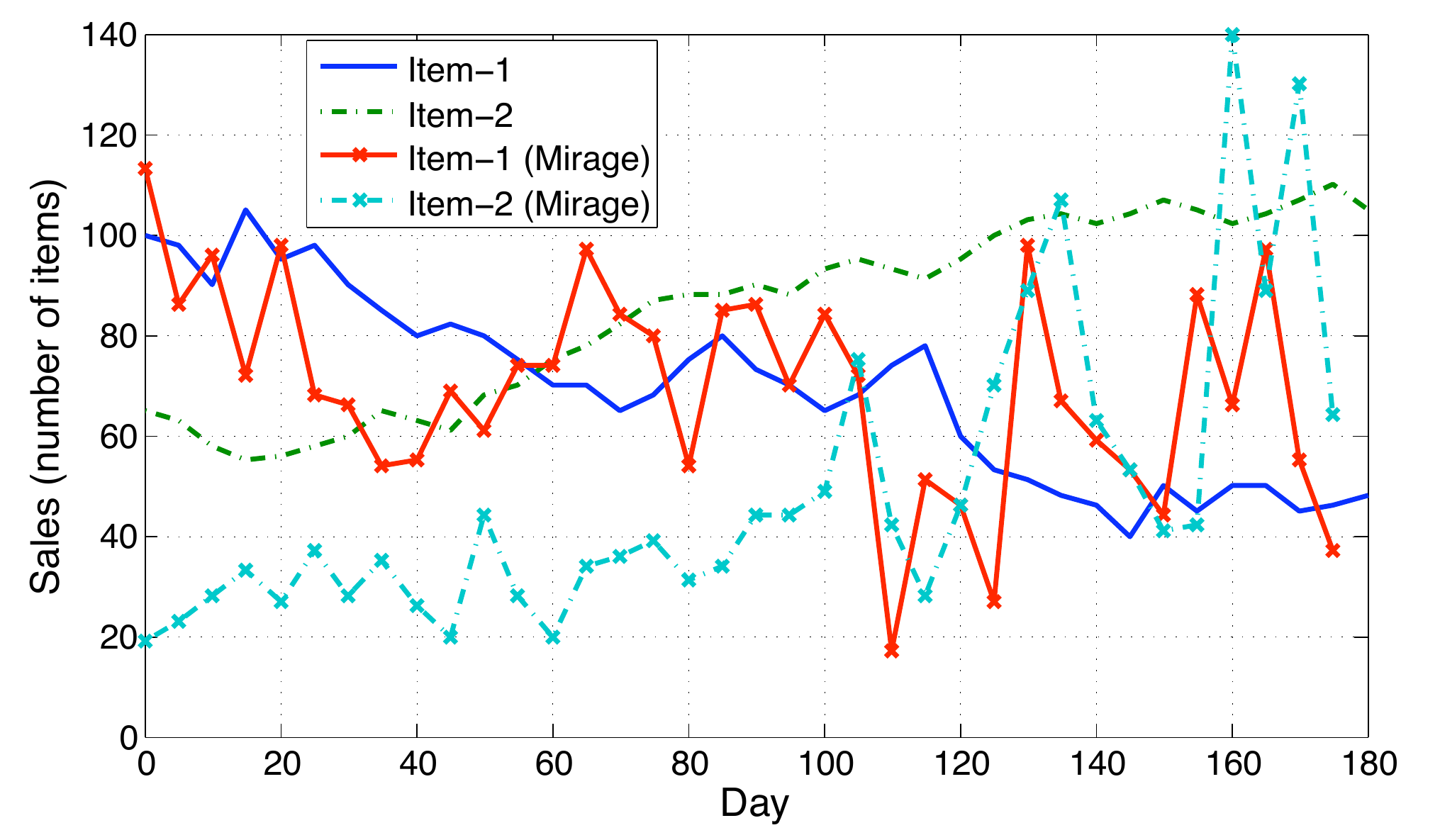} 
      \vspace*{5pt}
   \hrule
   \caption{The figure shows the results of the sales trend presented to an attacker by {\honeytoken}. The goal in this experiment was to obtain a random sales distribution. }
   \label{fig:salesrandom}
\end{figure}

\begin{figure}[t!]
   \centering
   \includegraphics[width=4in]{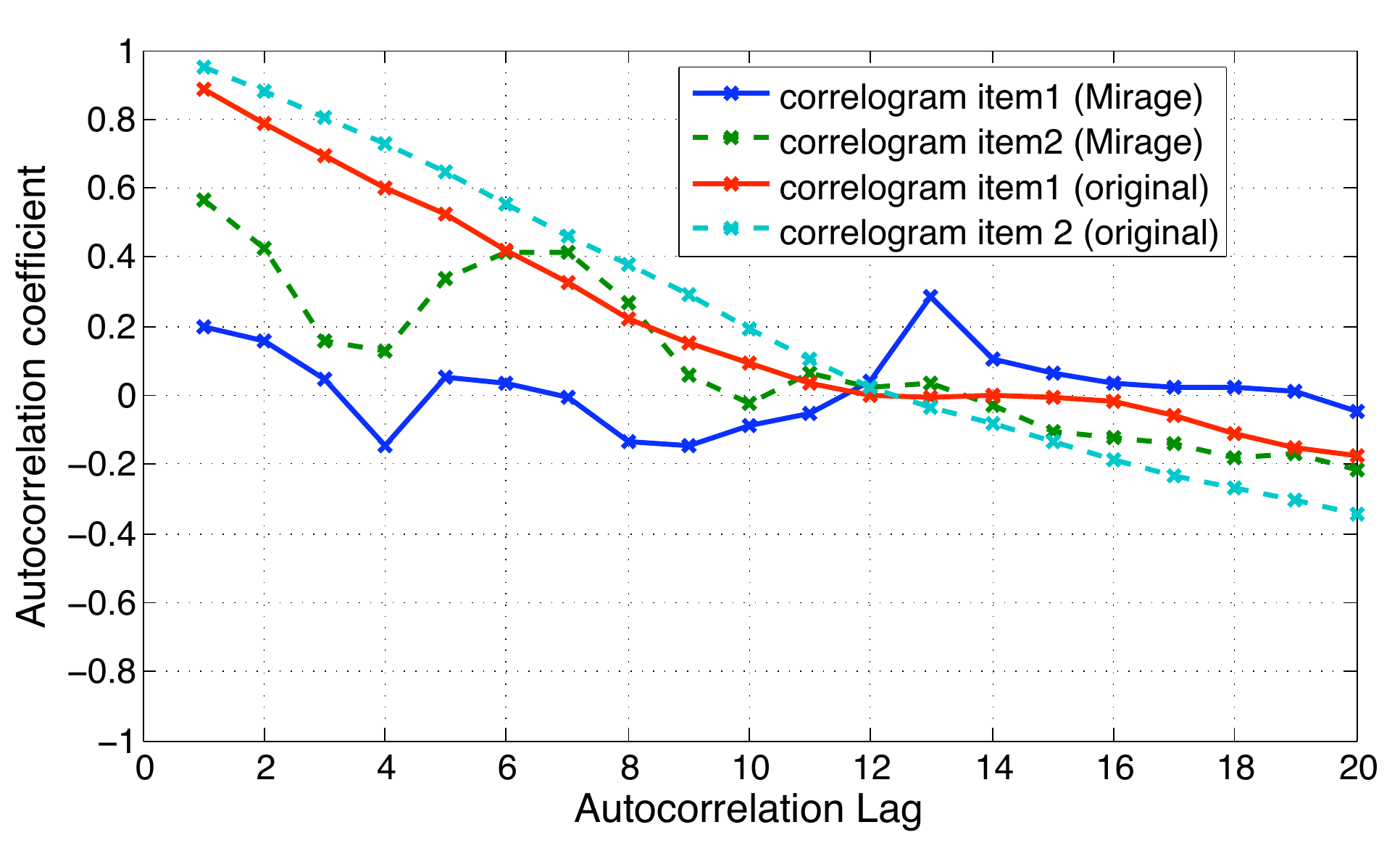} 
      \vspace*{5pt}
   \hrule
   \caption{The figure shows the correlogram  (autocorrelation at different lags) for original sales trend in Figure~\ref{fig:example3} and the correlogram after using {\honeytoken}.}
   \label{fig:corr}
\end{figure}

To mitigate the above problem, we set our goal to a random distribution. This goal is set to the maximum number of items on the shelf in the previous month and an additional random number between 0 and the total number of honeytokens. The results for the random-goal experiment is shown in Figure~\ref{fig:salesrandom} where 150 separate inventories were performed. To show whether the trends were near-random, we present a correlogram for the data in Figure~\ref{fig:corr}. A correlogram is an autocorrelation plot for different time lags. If data is truly random, the autocorrelation at different lags would be close to 0---we observe the autocorrelation coefficients are closer to zero for {\honeytoken} as compared to the original data. 


\subsection{Tradeoff with number of honeytokens}
\label{subsec:tradeoff}

\begin{figure}[t!]
   \centering
   \includegraphics[width=4in]{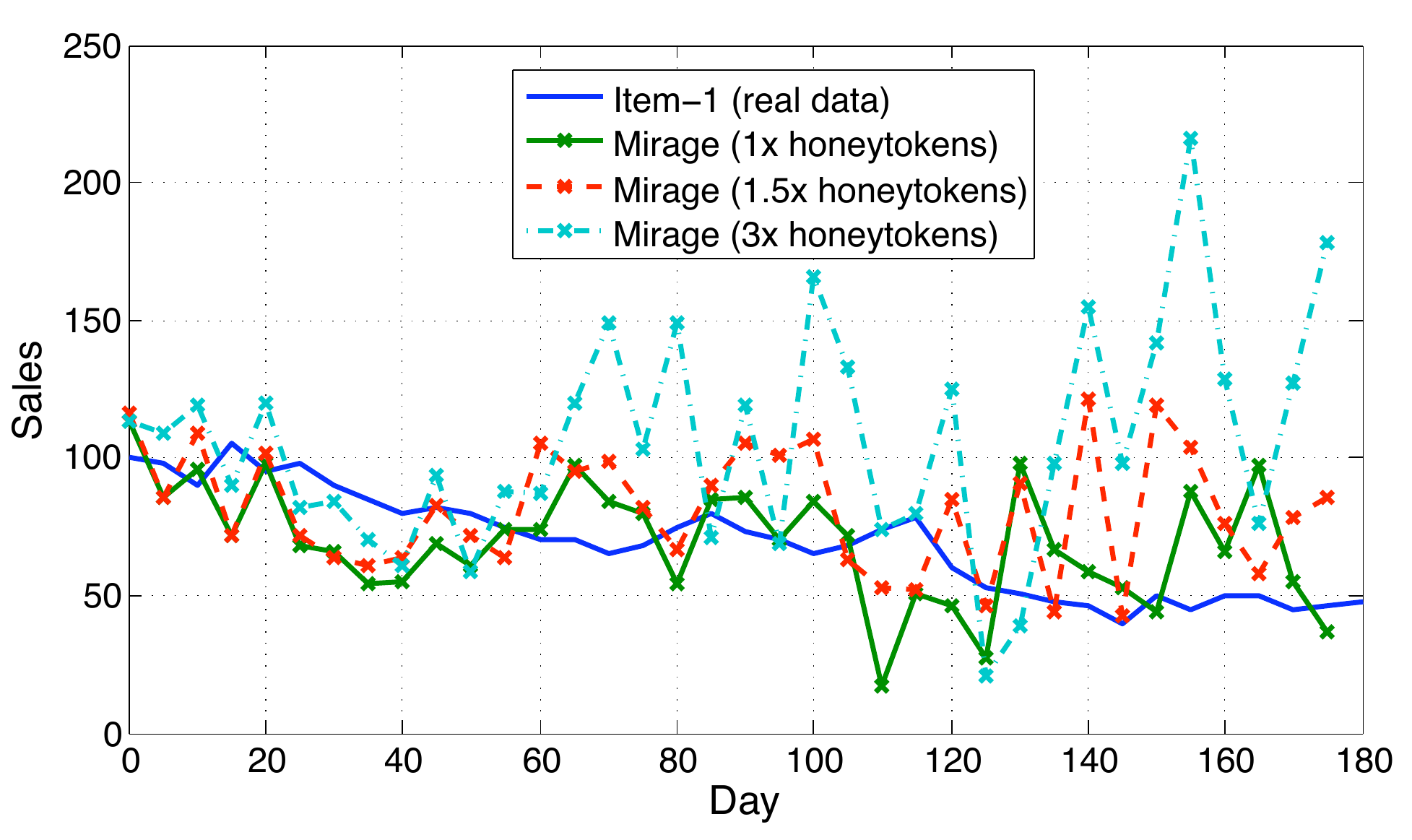} 
      \vspace*{5pt}
   \hrule
   \caption{The figure shows the sales trend seen by an attacker for {\honeytoken} with different number of honeytokens. This figure is only for item 1 in Figure~\ref{fig:example2}. }
   \label{fig:salesitem1}
\end{figure}

\begin{figure}[t!]
   \centering
   \includegraphics[width=4in]{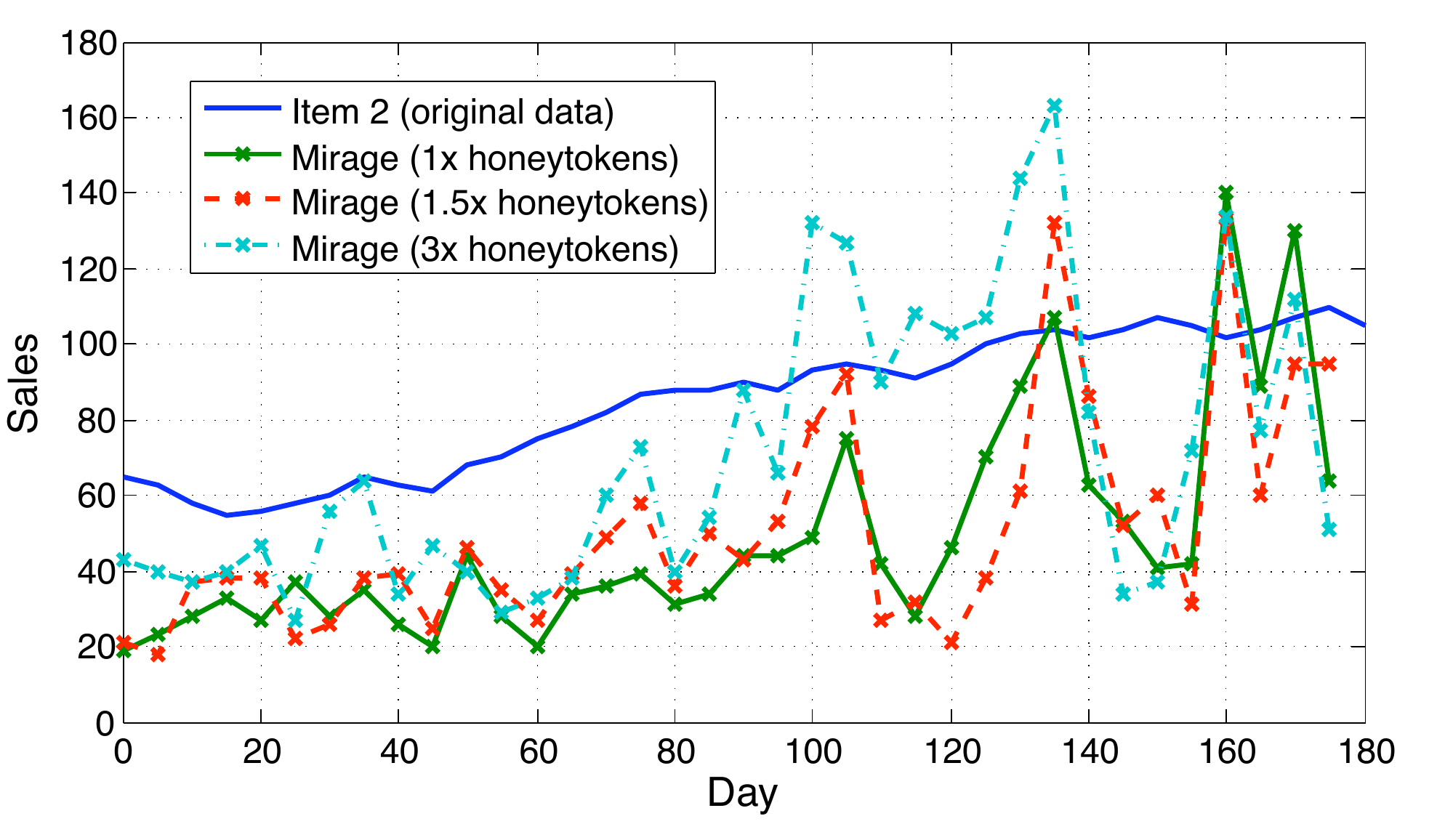} 
      \vspace*{5pt}
   \hrule
   \caption{The figure shows the sales trend seen by an attacker for {\honeytoken} with different number of honeytokens. This figure is only for item 2 in Figure~\ref{fig:example2}.  }
   \label{fig:salesitem2}
\end{figure}

\begin{figure}[t!]
   \centering
   \includegraphics[width=4in]{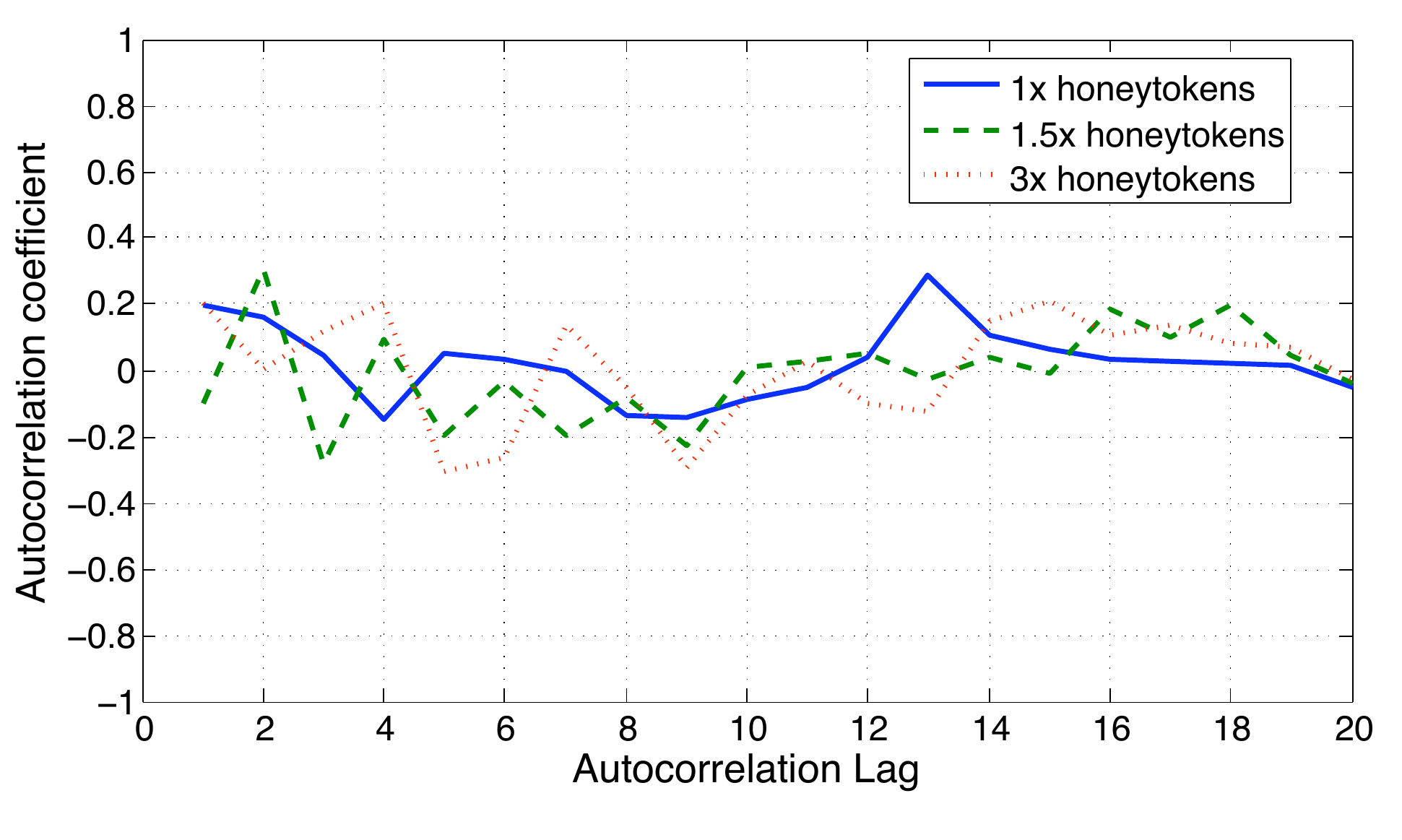} 
      \vspace*{5pt}
   \hrule
   \caption{The figure shows the correlogram (autocorrelation at different lags), for different number of honeytokens. As the number of honeytokens increases, the autocorrelation coefficient for different lags get closer to zero. }
   \label{fig:corrtradeoff}
\end{figure}

We next study the trade-off between the number of honeytokens and the degree of {\em randomness} produced by {\honeytoken}. Figure~\ref{fig:salesitem1} and Figure~\ref{fig:salesitem2} shows the trend that the attacker sees when the number of honeytokens is 100\%, 150\%, and 300\% of the total number of items. Note that our random goal in these experiments is a function of the number of honeytokens. Though other random goals are equally applicable, here we show that this simple function can produce desirable results in practice. Adding more honeytokens produces more {\em rises} and {\em falls} in the output trend---indication of better random output. However, this randomness comes at a cost of additional time to read the honeytokens. For example, for ``3$x$ honeytokens'', when over 400 tags were in use simultaneously for each inventory the read-success of scans reduced considerably.  For instance, on overage, 56\% of the tagged items were read, with a high of 84\% and a low of 15\% on particular inventories.  We anticipate that we would have obtained much better results had we scanned at a slower pace---effectively increasing the latency of inventorying. For our experiments, we walked up the warehouse aisles at the same speed regardless of the number of RFID tags on the shelf. Figure~\ref{fig:corrtradeoff} shows the correlogram for different number of honeytokens. Although the correlogram gets flatter with more honeytokens, the additional benefit diminishes. Therefore, using 1$x$ honeytokens presents a near optimal trade-off between randomness in the output data and overhead latency.


 \subsection{Overhead}
 \label{subsec:overhead}
 
\begin{figure*}[t!]
\begin{minipage}[t]{.5\textwidth}
\includegraphics[width=\textwidth]{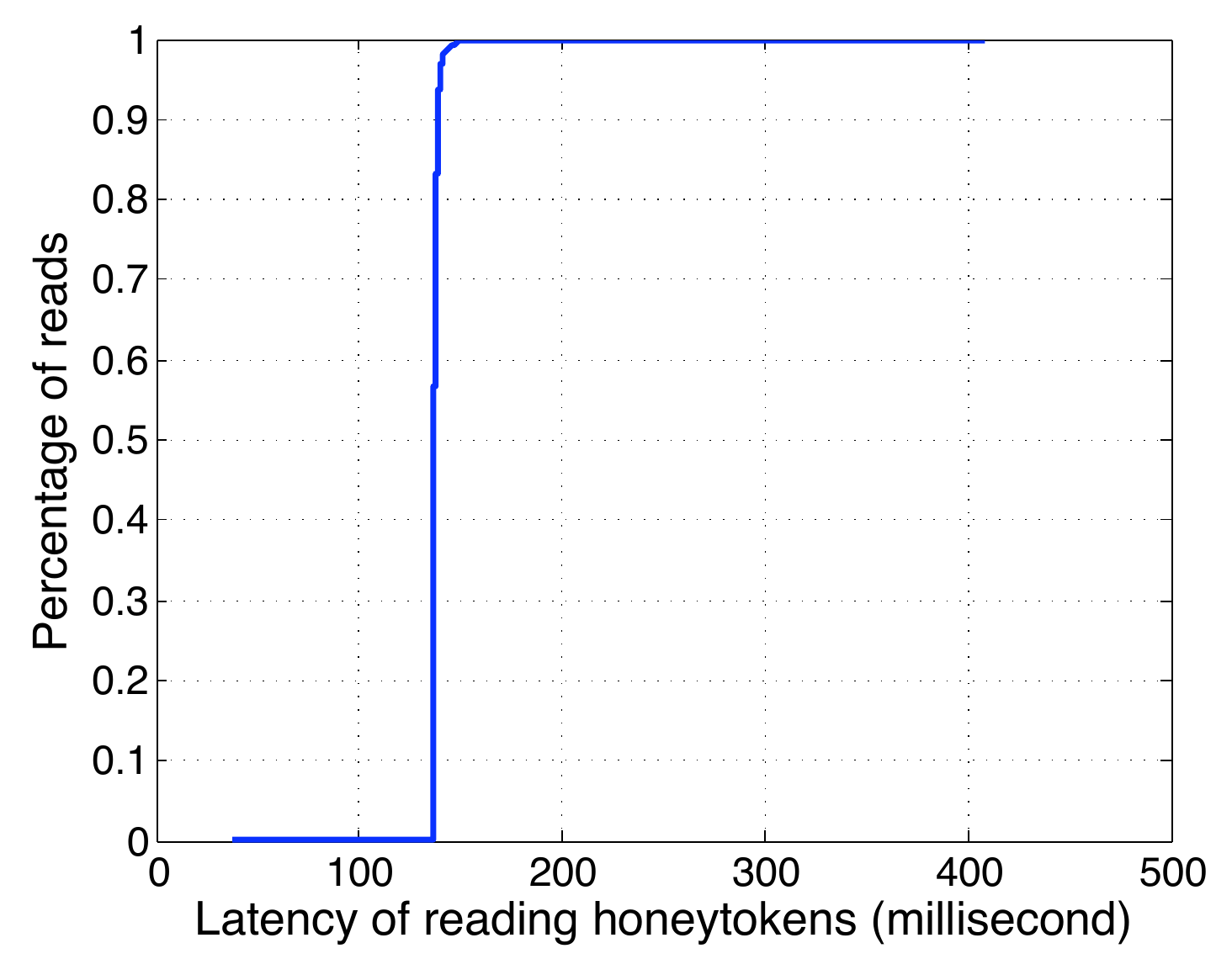}
   \vspace*{5pt}
   \hrule
\caption{The figure shows a cumulative probability distribution of the amount of time taken to read a honeytoken by our reader. The overhead associated with reading additional tags is small---less than 150 ms.}
\label{fig:read}
\end{minipage} \hspace{4pt}
\begin{minipage}[t]{.5\textwidth}
\includegraphics[width=\textwidth]{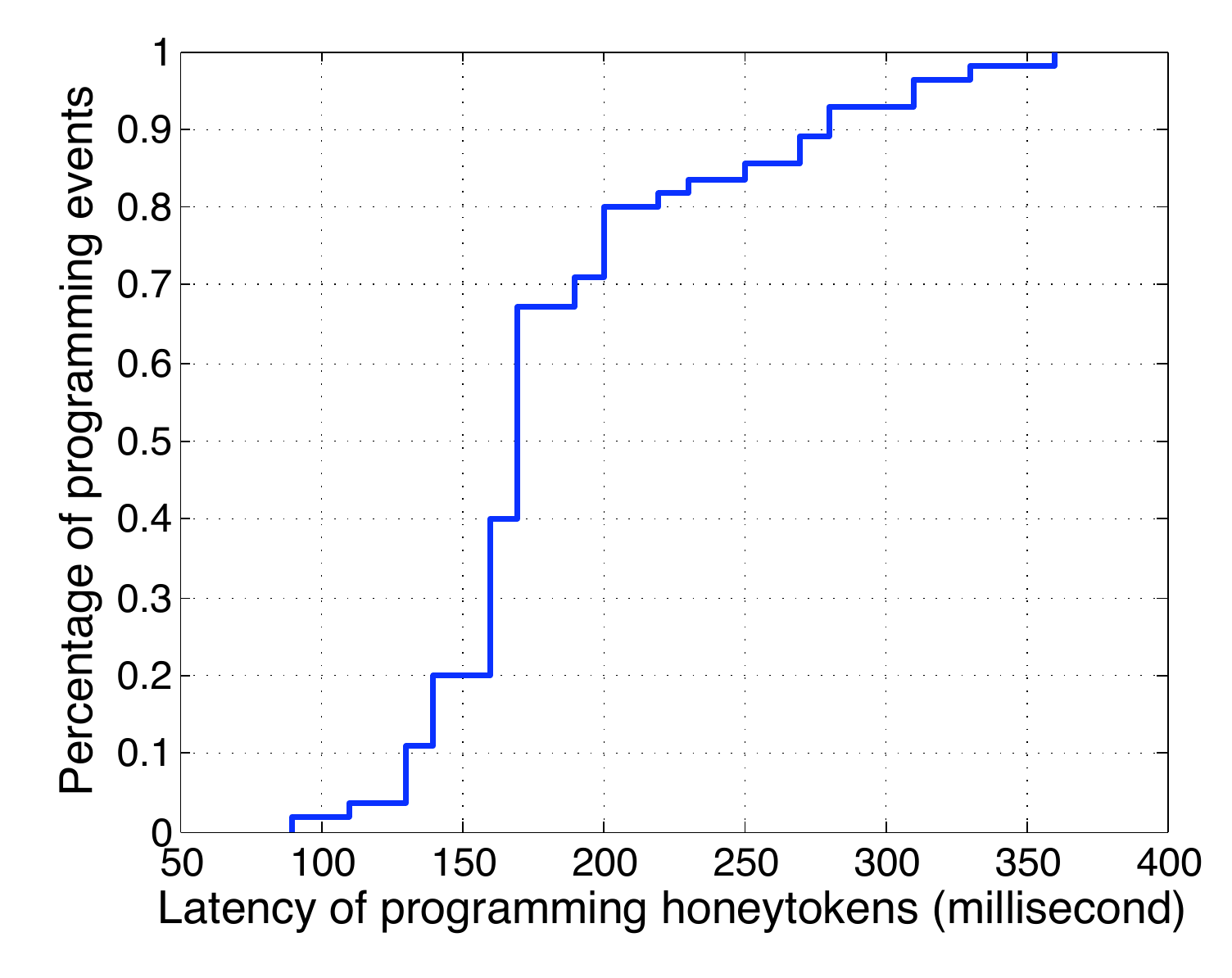}
   \vspace*{5pt}
   \hrule
\caption{The figure shows a cumulative probability distribution of the amount of time taken to reprogram a honeytoken tag (for activation and de-activation). This overhead varies from 90 ms to 350 ms.}
\label{fig:write} \end{minipage}
\end{figure*}

{\honeytoken} adds overhead to the inventory process in terms of additional tag-read and tag-programming (write) 
latency. If this latency is large per actual item read, retail store moguls such as Wal-Mart will eschew adopting {\honeytoken}. To quantify this overhead, we perform an evaluation of the additional read and write latency induced by {\honeytoken}. The read results were collected as part of the previous experiments and the write experiments were performed separately in a laboratory environment. Figure~\ref{fig:read} is a cumulative distribution function of the amount of time taken by our RFID reader to read  honeytoken tags and Figure~\ref{fig:write} is a CDF of the latency to reprogram a honeytoken tag.  Figure~\ref{fig:read} is calculated over 12,000 reads and Figure~\ref{fig:write} is calculated over 55 writes.  The read latency is always less than 150 ms and the write latency varies from 90 ms to 350 ms. Clearly the additional overhead induced by adding honeytokens is small as compared to solutions such as privacy zones~\cite{Rivest2003}. Therefore, {\honeytoken} can be used to de-trend sales, restocking, and inventory trend with minimal additional overhead. 

\section{Related Work}
\label{sec:rw}

Illicit inventorying is known to be a serious problem for RFID-enabled retail store environments. However, efforts to mitigate illicit inventorying is sparse in the literature---limited to a few research efforts. Here, we review previous literature that is most relevant and related to {\honeytoken}.

\subsection{Illicit Inventorying}
Two techniques have been proposed in the past for mitigating illicit inventorying. The first method~\cite{Syverson2005} uses an active device that can mimic several tags at once.  When an attacker performs an unauthorized inventory, several false ids are transmitted using the active device to the illicit reader.  The device houses a microcontroller that can execute logic to decide what to  transmit.  Such devices  are available for purchase~\cite{Rieback2005}, and they have been shown to operate as described.  However, the disadvantages of this solution include dollar cost, constant power to the active device, and conspicuous size that makes it easily noticeable to attackers.  Also, the devices require extensive upfront programming so that the identification values that are transmitted would not interfere with actual inventorying. A complimentary approach proposed is the idea of a blocker tag~\cite{Brainard}.  This is a specially designed tag that spams out several hundred random ids to all readers in range.  This tag could be carried by an individual when they do not want someone to illicitly scan the items they are carrying.  Though patented, such a device is not available for purchase. However, apart from the additional cost of designing such a tag (which is a huge deterrent for retail stores like Wal-Mart), the latency of reading an actual tag during a legitimate inventory could be huge---one tag broadcasts a large number of random IDs, potentially jamming the RF channel.

\subsection{Honeytokens usage}
 Also related to {\honeytoken} is the concept of honeytokens. Honeytokens have been used in personally identifiable information (PII) databases to detect unauthorized access~\cite{Mokube2007,McRae2007,Yuill2004}.  In industrial applications, privacy laws prohibit individuals from accessing personal information that they do not have a valid use for~\cite{McCormick2008}.  Detecting when unauthorized access has occurred is difficult, as the data requirements for applications and users vary greatly.  Honeytokens, which have no valid business use by definition as they are synthetic, are generated and then inserted into areas where users might be tempted to access them.  The honeytokens are monitored and if an individual accesses the record, alarms are generated.  The honeytokens are generated in large amounts so that the probability of a malicious user encountering them is increased.  Honeytokens work well in personal information databases as it is relatively easy to make large amounts of synthetic PII data that looks real and very valuable to an attacker~\cite{Valli2007}. However, in {\honeytoken}, we use honeytokens to add random noise to sales and restocking in retail store environment which is a completely different application domain.
 

\section{Conclusion}
\label{sec:concl}

In this paper, we design, implement, and evaluate {\honeytoken}, a system to mitigate the threat of illicit inventorying by an attacker in an environment that uses RFID tags.  {\honeytoken} uses a concept of honeytokens, additional programmable RFID tags, to inject random noise in a retail store environment. Through a history-based algorithm on the RFID reader, {\honeytoken} determines the number of honeytokens to activate or de-activate such that the attacker sees a random or flat sales or restocking trend. Through exhaustive evaluation in an actual experimental warehouse using an off-the-shelf RFID reader and inexpensive tags, we show that {\honeytoken} successfully de-trends sales and restocking events while adding minimal overhead to the inventorying process.

\bibliographystyle{abbrv}
\bibliography{mirage}

\end{document}